\documentclass[prb,twocolumn,floatfix,groupedaddress,showpacs]{revtex4}
\usepackage{graphicx}
\usepackage[dvips]{epsfig}

\newcommand{\be}{\begin{equation}}
\newcommand{\ee}{\end{equation}}
\newcommand{\ben}{\begin{eqnarray}}
\newcommand{\een}{\end{eqnarray}}

\newcommand{\nd}{\noindent}
\begin{document}

\title{The workings of the Maximum Entropy Principle in collective human behavior}

\author{A. Hernando$^1$, R. Hernando$^2$, A. Plastino$^{3,\,4}$, A. R. Plastino$^{3,\,5}$}
\affiliation{
$^1$ Laboratoire Collisions, Agr\'egats, R\'eactivit\'e, IRSAMC, Universit\'e Paul Sabatier
118 Route de Narbonne 31062 - Toulouse CEDEX 09, France\\
$^{2}$Social Thermodynamics Applied Research (SThAR),
Ambrosio Vallejo 16, 28039 Madrid, Spain\\
$^{3}$Instituto Carlos I de Fisica Teorica y Computacional
 and Departamento de Fisica Atomica, Molecular y Nuclear
 Universidad de Granada, Granada, Spain\\
$^4$National University La Plata, Physics Institute (IFLP-CCT-CONICET) C.C. 727, 1900 La
Plata, Argentina\\
$^5$CREG-National University La Plata-CONICET
 C.C. 727, 1900 La Plata, Argentina}

\begin{abstract}

\nd We exhibit compelling  evidence regarding how well does the
MaxEnt principle describe the rank-distribution of
city-populations via an exhaustive study of the 50 Spanish
provinces (more than 8000 cities) in a time-window of 15 years
(1996-2010). We show that the dynamics that governs the
population-growth  is the deciding factor that originates the
observed distributions. The connection between dynamics and
distributions is unravelled via MaxEnt.

\end{abstract}
\maketitle

\section{Introduction}

\nd Orderliness, reflected in either scaling properties
\cite{bat1} or power laws \cite{city1,power,pareto},  is
encountered in different frameworks involving social groups.
Salient example  is Zipf's law \cite{zipf}, a power law with
exponent $-2$ for the density distribution function that is
observed in describing urban agglomerations~\cite{ciudad} and firm
sizes all over the world~\cite{firms}. This kind of ``order" has
got immense attention in the literature. The pertinent
regularities have been found in other scenarios as well, ranging
from percolation theory and nuclear
multi-fragmentation~\cite{perco} to the abundances of genes in
various organisms and tissues~\cite{furu}, the frequency of words
in natural languages~\cite{zipf,zip}, the scientific collaboration
networks~\cite{cites}, the total number of cites of physics
journals~\cite{nosotrosZ}, the Internet traffic~\cite{net1} or the
Linux packages links~\cite{linux}. Another special
regularity in the density distribution function of the number of
votes in the Brazilian elections,\cite{elec1} a power law with exponent $-1$.
This kind of distribution is the origin of the so-called First-Digit Law, or Benford's Law\cite{benford}.
Such exponent has also been found in Ref.~\onlinecite{nosotros} by
recourse to an information-theoretic methodology\cite{alp}, for
both the city-size distribution of the province of Huelva (Spain)
and the results of the 2008 Spanish General Elections.
Also log-normal distributions are applied to cities' sizes.
While a city-size distribution is often
associated with Zipf's law, this holds only in the upper tail.
However the entire distribution of
cities, not just the largest ones, is close to log-normal (see Ref. \onlinecite{gibrat}).
Undoubtedly, some aspects of human behavior reflect a kind of ``universality".

\nd Common to all these disparate systems is the lack of a
characteristic size, length or frequency for the observable of
interest, which makes it scale-invariant. In Ref.~\onlinecite{nosotros}
we have introduced an information-theoretic technique based upon
the minimization of Fisher's information measure\cite{alp}
(abbreviated as MFI) that allows for the formulation of a
``thermodynamics" for scale-invariant systems. The methodology
 establishes an analogy between such systems  and
physical gases which, in turn, shows that the two special power
laws mentioned in the preceding paragraph lead to a set of
relationships formally identical to those pertaining to the
equilibrium states of a scale invariant non-interacting system,
the \emph{scale-free ideal gas} (SFIG). The difference between the
two  distributions is thereby attributed  to different boundary
conditions on the SFIG.

\nd Many social systems, however, are not easily included into the
two types of description of the preceding paragraph. Inspired in
opinion dynamics models\cite{oppi,elec1,elec2} we described in
Ref. \onlinecite{oursnow3} a numerical process that reproduces the
shapes of the empirical city-population distributions. The model
is based in a competitive cluster growth process inside an
scale-free ideal network (a scale-free network which degree
distribution is described as an SFIG). The finite size of the
network introduces competition, and thusly correlations in the
cluster sizes. The larger the competitiveness, the larger the
deviation from the SFIG distribution. The equivalence with the
workings of physical gases is compelling indeed. However, some
aspects of the concomitant problems defy full understanding, since
an analytical prescription for the classification of the
size-distribution of social groups was missing. {\it In order to
do look for such analytical understanding we will appeal here to a
variational approach centered on information theory.}

\nd We have recently shown (see Ref. \onlinecite{XXX}) how to deal
with the well-established MaxEnt technique \cite{jaynes,katz} by
including, within the associated Lagrangian, information regarding
equations of motion, i.e., ``dynamical" information of a kind that
goes beyond the customary one, based upon expectation values. We
demonstrated that proportional growth and hyper-exponential growth
can both be described in dynamical terms, accurately predicting
the equilibrium distribution for such systems. This is tantamount
to  giving a dynamical interpretation to the `information cost'
introduced in Ref. \onlinecite{X1}. Our  main present purpose is
to apply what we  learned in \onlinecite{XXX} to empirical
systems,  this demonstrating i) the applicability of MaxEnt to
collective human behavior, and ii) the potential use of an
explicit \emph{social thermodynamics}.

\nd This work is is organized as follows: in Sec. II we revisit
the theoretical MaxEnt approach to be followed here. In Sec. III
we present  empirical data for the Spanish provinces. The ensuing
 results clearly show the usefulness of the theoretical method. In
Sec. IV we discuss some noteworthy features of the results an
their possible uses in other contexts. A summary is given in Sec.
V and, finally,  some technical aspects are the subject of  the
Appendix.

\section{Theoretical approach}

\subsection{Basics}

\nd Following the methodology described in \onlinecite{XXX}, let us
define
\begin{description}
  \item[i)]   $N$ as the total population,
  \item[ii)]  $n_c$ as the total number of cities into which people are apportioned,
  \item[iii)] $x_i$ as the population of the $i$-th city,
  \item[iv)]  $x_0$ as the minimum possible population amount (which is at least $x_0=1$),
  \item[v)]   $p(x)$ as the relative number of cities with exactly a population amount given by $x$.
\end{description}
\nd Considering the continuous limit of the distribution
$p(x)$, the conservation of both $N$ and $n_c$ guarantees
\begin{description}
  \item[i)]  $\displaystyle \int_{x_0}^\infty dx p(x) = 1$, and
  \item[ii)] $\displaystyle \int_{x_0}^\infty dx p(x)x = N/n_c$.
\end{description}
We now turn on  MaxEnt: in dynamical equilibrium, the distribution
$p(x)dx$ is the one that maximizes  Shannon's entropy $S$ subject
to the above  constraints. One
 thereby concocts  the
variational Lagrangian
\begin{equation}
H = S -\mu n_c - \lambda N,
\end{equation}
with $\mu$ and $\lambda$ the corresponding Lagrange multipliers.
MaxEnt entails $\delta H = 0$.

\subsection{Dynamics}

\nd We showed in Ref. \onlinecite{oursnow3} that  cities' population
expansion can be modelled by means of clusters' growth in complex
networks. Such process (diffusion) in networks generally starts i)
using a node as a seed, ii) its first neighbors being added to the
cluster in the first iteration, iii)  the neighbors of those
neighbors afterwards, and so on. Mathematically, if $\Delta t$ is
the interval of time for each iteration and $x(t)$ is the
population of the cluster at the time $t$, we can write
\begin{equation}
x_i(t+\Delta t) = x_i(t)+\Delta t \sum_{j=1}^{x_i(t)} k_{j}
\end{equation}
where $k_{j}$ is the number of neighbors of the $j$-th node
accepted to the cluster per unit time. According to the central
limit theorem, we generalize this equation as
\begin{equation}\label{clt}
x_i(t+\Delta t) \simeq x_i(t) + \Delta t\left(k x_i(t) \pm \sigma_k\sqrt{x_i(t)}\right),
\end{equation}
where $k$ is the global mean value of accepted neighbors per unit
time and $\sigma_k$ its standard deviation. For large enough values
of the population this last term can be neglected, and write for the
continuous limit
\begin{equation}\label{pgrow}
\dot{x}_i(t) = k x_i(t),
\end{equation}
where the dot represents the time derivative. This equation
represent the so-called proportional growth, whose generalization
to a random process leads to the so-called geometrical Brownian
motion. We show in Ref. \onlinecite{XXX} how to proceed to
construct the entropy $S$ for such a scale-free dynamics.

\subsection{MaxEnt for scale invariant systems}

\nd  We now consider a system governed by the proportional growth
of Eq. (\ref{pgrow}), with $k$ being regarded as reflecting a
Wiener process. Our equation is linearized by introduction of the
new variable $u=\log(x/x_0)$, which leads to $\dot{u}=k$, and then
to a Shannon measure expressed in terms of $u$
\begin{equation}
S = \int_0^\infty p(u)\log(p(u))du.
\end{equation}
With our two  conservation rules written now in  $u-$terms, the
MaxEnt variational problem becomes
\begin{equation}
\delta H = \delta\left[ S - \mu\int_0^\infty du p(u) -
\Lambda\int_0^\infty du p(u) e^u \right]=0,
\end{equation}
where we have employed the definition $\Lambda=x_0\lambda$. The
solution is the equilibrium density $p(u)du = Z^{-1}\exp(-\Lambda
e^u)du$, that one now recasts in terms of the observable $x$
getting
\begin{equation}\label{distri1}
p(x)dx = Z^{-1}\frac{e^{-\Lambda x/x_0}}{x}dx.
\end{equation}
From the conservation rules we easily obtain for the pertinent
constraints the values
\begin{equation}\label{lamb}
\begin{array}{c}
Z = \Gamma(0,\Lambda)\\
\displaystyle \frac{e^{-\Lambda}}{\Lambda\Gamma(0,\Lambda)} =
\frac{N}{n_c x_0},
\end{array}
\end{equation}
with $\Gamma(a,z)$ the so-called incomplete Gamma function.

\subsection{The Gamma Scaling Law}

\nd The concomitant cumulative function $P(x)$ reads
\begin{equation}
P(x) = 1-\frac{\Gamma(0,\Lambda x/x_0)}{\Gamma(0,\Lambda)},
\end{equation}
and the associated  rank-distribution (RD) ---obtained from the
inversion of the cumulative one--- becomes
\begin{equation}\label{RD1}
x =
\frac{x_0}{\Lambda}\Gamma^{-1}\left[0,\Gamma(0,\Lambda)r/n_c\right],
\end{equation}
where $r$ is the (continuous) rank from 0 to $n_c$, and
$\Gamma^{-1}(z)$ denotes the inverse function of $\Gamma:$
$\Gamma(\Gamma^{-1}(z))=z$. We derive from here what we call
the \emph{Gamma Scaling Law}
\begin{equation}
\begin{array}{c}
\displaystyle x' = x\Lambda/x_0\\
\displaystyle r' = r \Gamma(0,\Lambda)/n_c,
\end{array}
\end{equation}
obtaining a  ``scaled" RD
\begin{equation}
x' = \Gamma^{-1}\left(0,r'\right),
\end{equation}
which no longer depends on $N$, $n_c$, or $x_0$.

\subsection{Beyond  proportional growth}

\nd We now consider the more general expression
\begin{equation}
\dot{x}(t)=kx^q(t),
\end{equation}
where $q$ is a dimensionless parameter. Linearization is achieved
by means of the q-logarithm of Tsallis' statistics\cite{tsallis}
\ben \label{qlog} & \log_q(x)= \frac{x^{1-q} - 1}{1-q}; \,\,\,[x > 0; \log_1(x)= \ln{x}],
 \cr &
  e_q(x)=[1+(1-q)x]_+^{1/(1-q)},\een
defining the new variable $w=\log_q(x/x_0)$, and so obtaining $\dot{w}=k$.
The ``$w$"-Shannon entropy becomes
\begin{equation}
S = \int_0^\infty p(w)\log(p(w))dw.
\end{equation}
The concomitant MaxEnt problem is of the form
\begin{equation}
\delta H = \delta\left[ S - \mu\int_0^\infty du p(w)-\Lambda\int_0^\infty du p(w) e_q(w) \right]=0,
\end{equation}
whose solution is $p(w)dw = Z^{-1}\exp(-\Lambda e_q(w))dw$. Changing
back to the observable $x$ we find
\begin{equation}
p(x)dx = Z^{-1}\frac{e^{-\Lambda x/x_0}}{x^q}dx.
\end{equation}
The constraints derive from the conservation rules in the usual
manner as
\begin{equation}
\begin{array}{c}
Z = (\Lambda/x_0)^{q-1} \Gamma(1-q,\Lambda)\\
\displaystyle\frac{\Gamma(2-q,\Lambda)}{\Lambda\Gamma(1-q,\Lambda)} = \frac{N}{n_c x_0}.
\end{array}
\end{equation}
The current cumulative function turns out to be
\begin{equation}
P(x) = 1-\frac{\Gamma(1-q,\Lambda x/x_0)}{\Gamma(1-q,\Lambda)},
\end{equation}
and for the associated RD has
\begin{equation}\label{RDq}
x =
\frac{x_0}{\Lambda}\Gamma^{-1}\left[1-q,\Gamma(1-q,\Lambda)r/n_c\right].
\end{equation}

\subsection{Proportional drift in q-exponential growth}

\nd  We now consider a $q$-equilibrium system subject to a slow
proportional drift, that may account for the ``natural"
population-growth or fluctuations of proportional nature.
One is here including a kind of noise that
affects the underlying dynamical equation via
\begin{equation}
\dot{x}=k_1x+k_qx^q,
\end{equation}
or, in terms of $u=\log(x)$,
\begin{equation}\label{dynqs}
\dot{u}=k_1+k_qe^{(q-1)u},
\end{equation}
where $k_1$ characterizes  proportional drift and $k_q$
hyper-exponential growth. Considering $k_1$ also as a Wienner
process, the system departs equilibrium  via proportional
diffusion. As seen in Ref. \onlinecite{XXX}, the kernel function for that kind
of diffusion is a log-normal distribution. We face the convolution
\begin{equation}\label{distriqs}
\begin{array}{l}
p(x;\Lambda,x_0,q,\sigma)dx =\\
\displaystyle\frac{dx}{(\Lambda/x_0)^{q-1}\Gamma(1-q,\Lambda)}
\int_{x_0}^\infty dx'\frac{e^{-\Lambda x'/x_0}}{x'^q}
\frac{e^{-\log^2(x'/x)/2\sigma^2}}{x\sqrt{2\pi}\sigma},
\end{array}
\end{equation}
where $\sigma$ is the drift-scale parameter. The mean value reads
\begin{equation}\label{eos}
\frac{\Gamma(2-q,\Lambda)}{\Lambda\Gamma(1-q,\Lambda)}
e^{\sigma^2/2}= \frac{N}{n_c x_0}.
\end{equation}

\section{Empirical observations}

\subsection{Description of the data used in this work}

\nd     The raw data are obtained from the Spanish state institute
INE and cover annually the period 1996-2010 (with the exception of
1997). It encompasses up to 8000 municipalities (the smallest
Spanish administrative unit) distributed within 50 provinces (the
building blocks of the autonomous communities, equivalent of the
USA-states). The autonomous cities of Ceuta and Melilla are not
included. We use provinces and municipalities as the closest
representatives of the ideal of a closed system's  fundamental
elements. But some arbitrariness remains in the data, since i)
there are many municipalities that actually contain more than one
population-center, ii) provinces that contains more than a single
socioeconomic cluster, and iii) economic regions that extend
beyond the frontiers of a province. Those facts introduce systematic
errors into the data, but we retain enough accuracy for the
purposes of this work.

\nd    The appropriateness of provinces as the proper scale for
the analysis of this work is due to a trade-off between large
enough data sets and locality. Autonomous communities as a whole
provide a large enough  sample, but the statistical properties we
are interested in change within them.  Smaller administrative
units called {\it comarcas} have a more local nature, but only
contains a
 few municipalities and thus  introduce large numerical errors. We
consider that choosing  the adequate scale is the most important
ingredient for an study of the kind we attempt here.

\nd    The data cover a total Spanish population of 39106917
inhabitants in 1996, reaching 47254510 in 2010. For this last
year, the largest municipality (Madrid) covers 3273049 people and
the smallest one just 5  persons (Ill\'an de Vacas, Toledo). The
total population of each province, $N$, ranges between 6458684
inhabitants (Madrid) and 95258 (Soria), and the number of
municipalities, $n_c$, between 371 (Burgos) and 34 (Las Palmas).
These
 $n_c-$figures are not large enough to build up an accurate density
distribution. Accordingly, we have  systematically worked with RD
instead. RD are usually built from the vector
$\mathbf{x}=\{x_i\}_{i=1}^{n_c}$, where $x_i$ is the population of
the $i$-th municipality. We assign rank numbers ranging from 1 to
$n_c$, from the largest ($\max(\mathbf{x})\rightarrow r=1$) to the
lowest (in population) ($\min(\mathbf{x})\rightarrow r=n_c$). In
order to compare with theoretical, continuous rank-distributions,
we have found it to be more accurate to assign ``middle-point"
values from $r=0.5$ to $r=nc-0.5$, instead.

\nd    For our analysis we generate, for each province and for
each time-period, the pair $(u_i,\dot{u}_i)$, with the logarithm
of the population $u_i$ correlated with the relative
population-increment $\dot{u}_i$. They are  computed as
\begin{equation}
\begin{array}{l}
\displaystyle u_i = \log[x_i(t_2)x_i(t_1)]/2\\
\displaystyle \dot{u}_i = \frac{\log[x_i(t_2)/x_i(t_1)]}{t_2-t_1},
\end{array}
\end{equation}
where $t_1$ and $t_2$ are consecutive years for which data are available.

\subsection{Zeroth order approximation: proportional growth}

\subsubsection{A paradigmatic example: Alicante}

\begin{figure}[t]
\begin{center}
\includegraphics[width=0.45\textwidth]{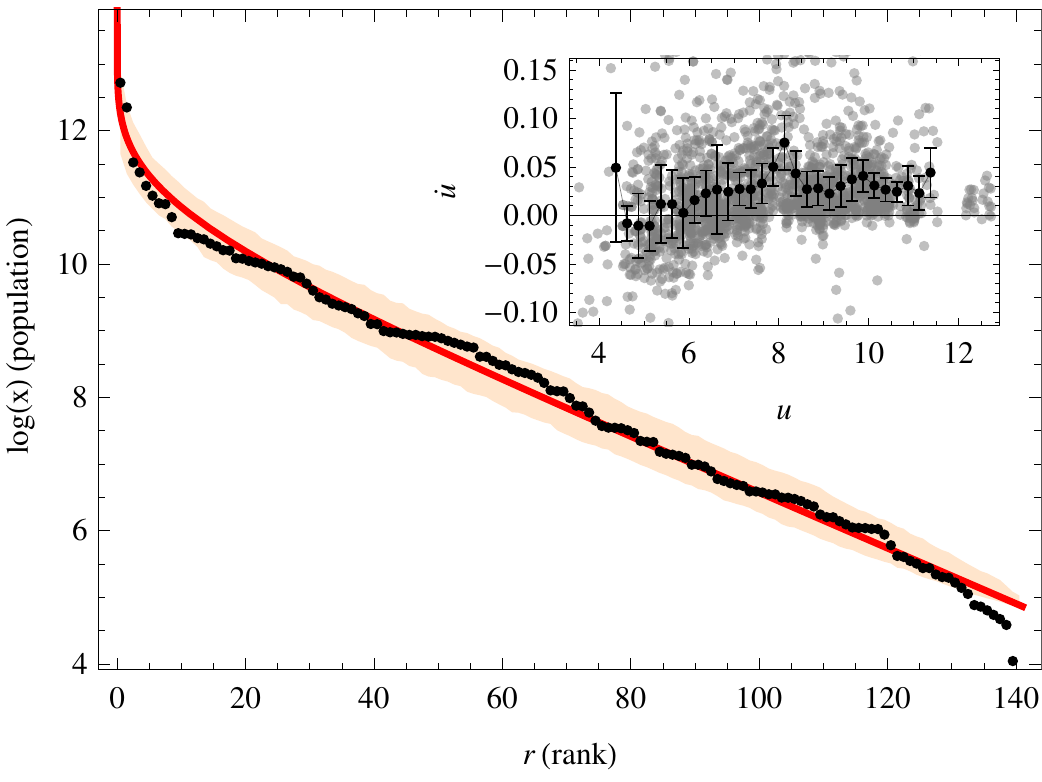}
\includegraphics[width=0.45\textwidth]{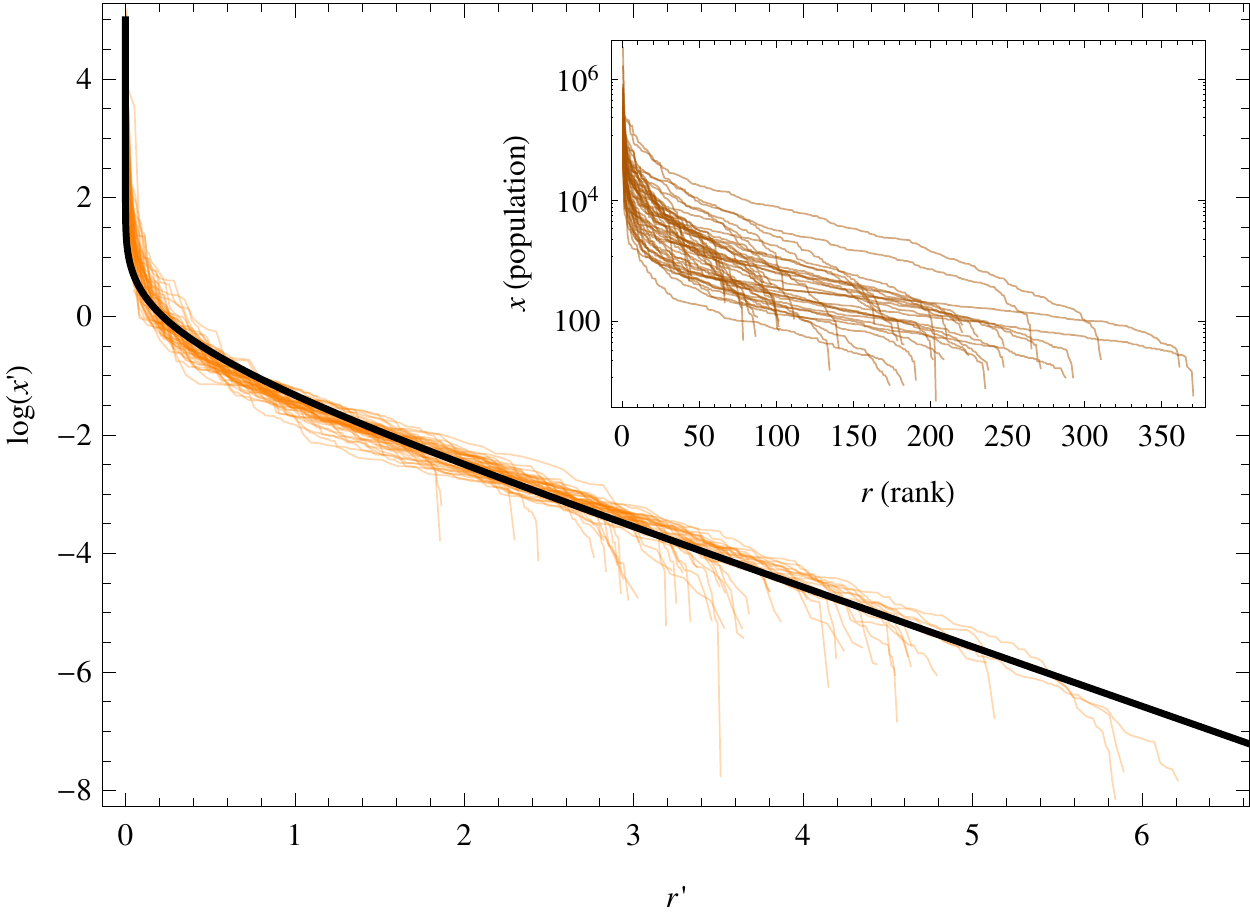}
\caption[]{Top panel: Rank-distribution of the province of Alicante (2010),
compared with the MaxEnt prediction for proportional growth using
the actual total population $N=1926285$ and the estimation
$x_0=125$. The shadowed area represents the 90\% confidence
interval for $n_c=141$ (2010). Inset: relative change vs. natural
logarithm of the population for this province for the last 15
years, where each gray dot represents one municipality for each
annual period. Black dots denote the mean value in each interval
$\Delta u=1$ and the error bars the standard deviation at that
interval. Bottom panel: Gamma-scaled rank-distribution of all 50 provinces.
Inset: raw data (2010).}\label{fig1}
\end{center}
\end{figure}

\nd    We use as paradigmatic example the province of Alicante to
illustrate our empirical analysis. The inset in the top panel of
Fig.~\ref{fig1} depicts $u$ versus $\dot{u}$ for all our
municipalities in the last 15 years. The correlation coefficient
between the two variables is found to be $0.16$, small enough to
consider, at zeroth order approximation, that the growth is of
proportional nature, independently of the size, thus obeying
Eq.~(\ref{pgrow}).

\nd    As seen in the previous section, the equilibrium
rank-density predicted for proportional growth by  MaxEnt follows
Eq.~(\ref{distri1}), where $\Lambda$ is univocally determined by
Eq. (\ref{lamb}) with i) an estimated minimum size $x_0$, ii) the
given total population $N=1926285$ (in 2010) and, iii) the number
of municipalities $n_c=141$.  $N$ and $n_c$ are well determined in
all cases. In assigning  $x_0$,
 fluctuations have an important influence. The easier way of estimating it is
 via
extrapolation (in the rank-plot) of a linear fit to the logarithm
of the lowest populations. A more sophisticated method employs a
non-linear fitting of the raw data to the MaxEnt rank-distribution
via this single parameter, using Eq. (\ref{RD1}) together with the
definition of $\Lambda$ Eq. (\ref{lamb}). Such simple procedure in
the log-scale works nicely enough. Note that $\log(x_0)$ is just a
shift of the distribution as a whole, playing no role in its
actual shape. We have obtained $\log(x_0)=4.826$ ($x_0=125$), with
an standard error of 0.036 and a correlation coefficient  of
$R=0.99968$. The comparison of the raw data with the analytical
rank-distribution is displayed in the top panel of
Fig.~\ref{fig1}. The shadowed area represents the 90\% confidence
interval due to finite size effects for $n_c=141$, numerically
estimated, as described in the Appendix. Since the empirical
rank-distribution falls inside the confidence interval, and the
observed dynamics are compatible with the dynamical assumption of
Eq. (\ref{pgrow}), we consider that this case constitutes  strong
evidence for the applicability of the MaxEnt principle.

\subsubsection{More detailed analysis}

\nd    We have applied the same (zeroth order) approximation of
proportional growth to the rest of Spain's provinces, also
including the values of the total populations $N$ in the fitting
procedure. Using the corresponding values of $n_c$ together with
the estimated values of $x_0$ and $N$ we get the value for
$\Lambda$, and apply then the Gamma-scaling law described in the
previous Section. The scaled rank-distributions are displayed in
Fig.~2 (raw data in the inset), and the numerical values of the
fitted results can be found in the Additional Material\cite{AM}. A quite
nice adjustment ensues in general (correlation coefficient of
$R=0.99984$ for Tenerife)  but not in all instances ($R=0.99635$
for Guadalajara). This ``failure" is linked to the strong
correlations between $\dot{u}$ and $u$ for some provinces, that
reaches significant values (of up to $R=0.57$  for the province of
Lugo). Although such correlations  may compromise the validity of
the zeroth order approximation of proportional growth, the quite
nice scaling-features exhibited by this plot are remarkable
indeed.

\nd    It is worth mentioning that the fitted value for the total
population is found to be systematically lower than the actual
value for it. Such scenario usually changes when the capital city
of the province is not considered in the fitting process, and thus
in the estimation of $N$. This indicates that the population of
the capital city is systematically larger than what one would
expect from the MaxEnt prediction (off  the 90\% confidence
level). This is not surprising since, as mentioned above,
provinces are not ideal, isolated systems: the actual
administrative municipality for capital cities are usually the sum
of the historical ones plus some near neighbors, and their economy
is expected to be highly correlated with that of other capital
cities. Although only 50 cities is a small sample, we have studied
the rank-distribution of these capitals and their dynamics to shed
some light onto this observation. We have found the pleasing
result that these capitals form {\it a scale-free system of their
own}. Comparing the rank-distribution with the MaxEnt prediction
that uses i) the actual value $N=15528025$ with ii) the only
fitting parameter $\log(x_0)=10.80\pm0.05$ ($R=0.99989$, higher
than that of Alicante) and encounter that  all the 50 cities get
located  inside the confidence level. When studying the relative
increment $\dot{u}$ vs. the log-size $u$, we find a very low
size-dependencies ($R=0.15$, lower than the one prevailing for
Alicante). This observation confirms the appropriateness of a
proportional growth dynamics. Again, the capital city of Madrid
exhibits a  larger-than-expected population (just in the limit of
the 90\% confidence level), indicating the possible existence of a
higher, international system of correlations.

\begin{figure*}[ht!]
\begin{center}
\includegraphics[width=0.31\textwidth]{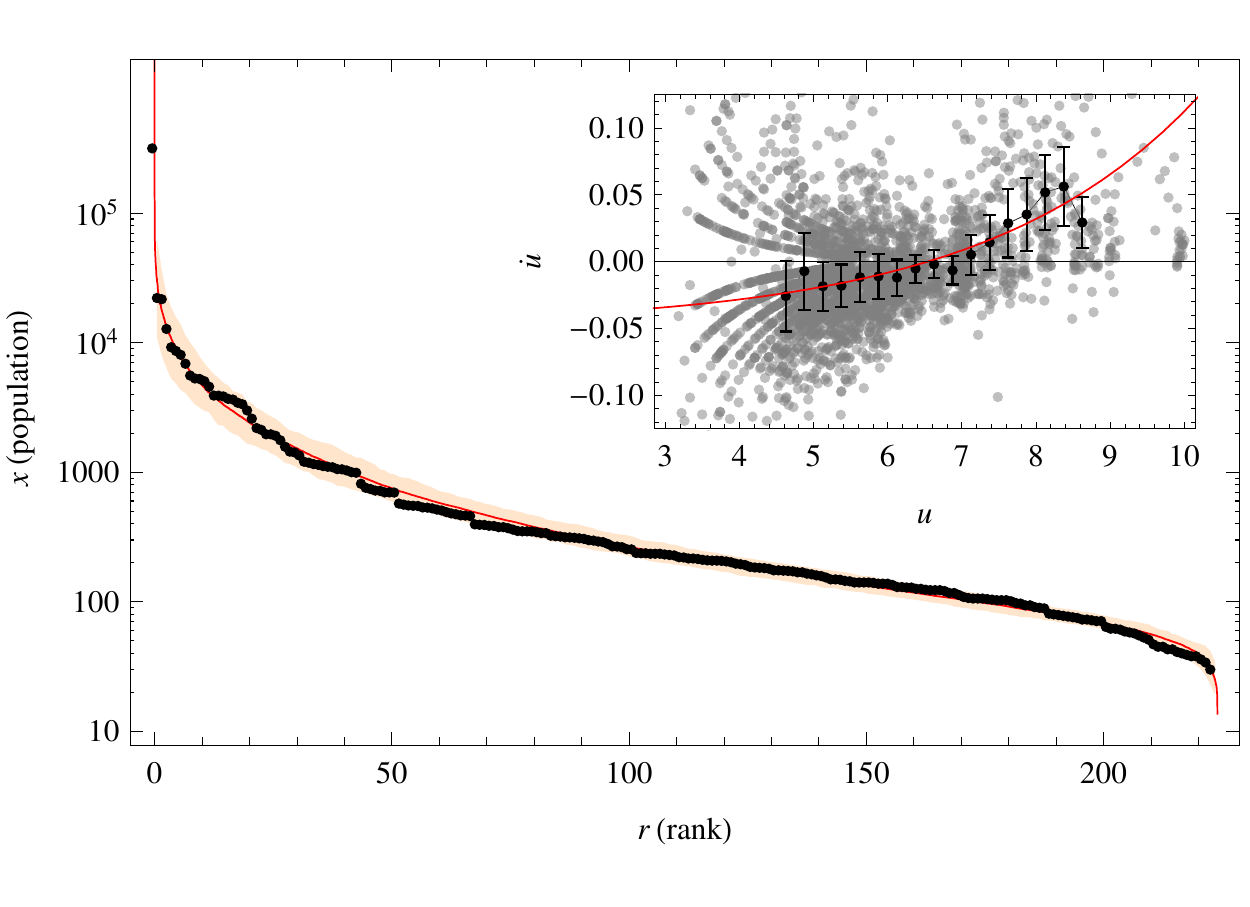}
\includegraphics[width=0.31\textwidth]{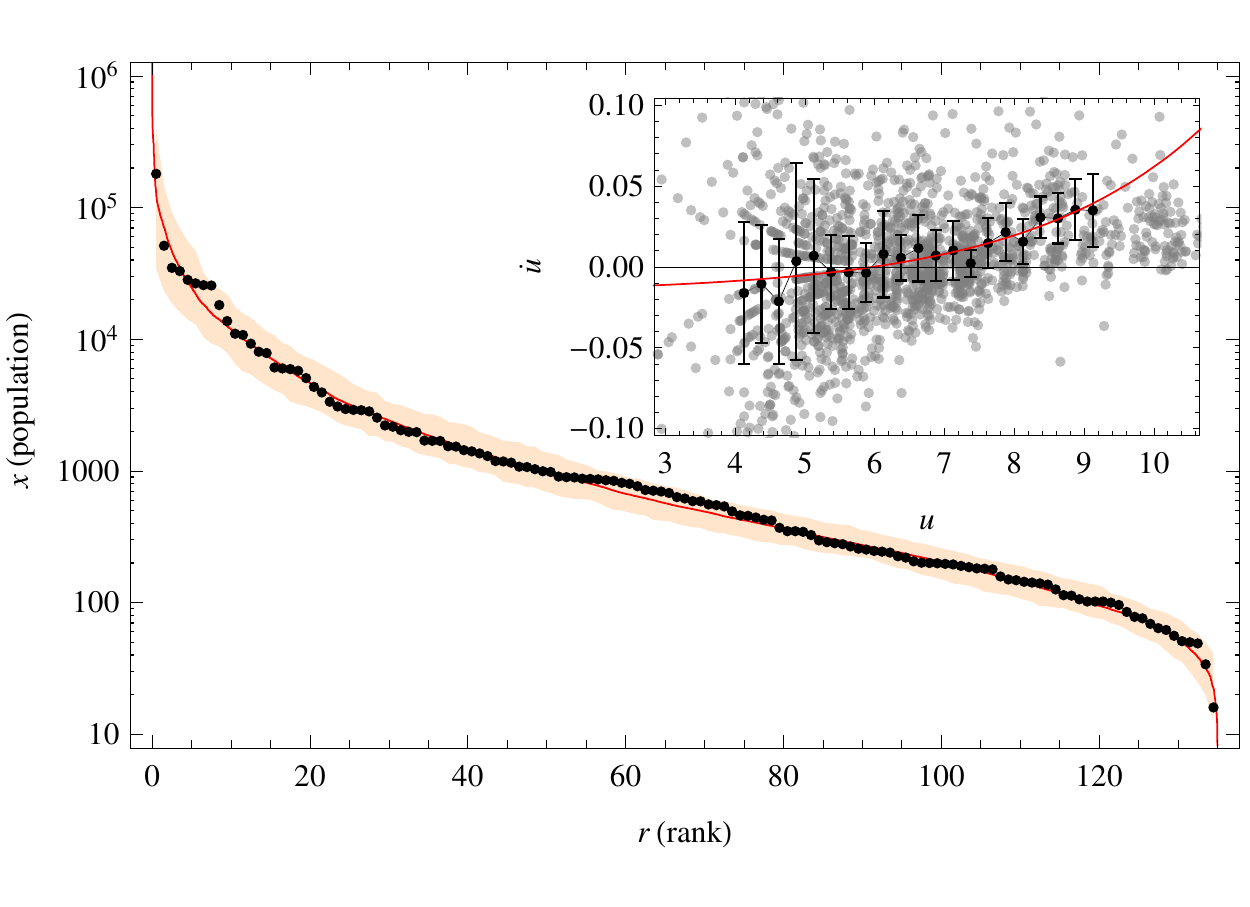}
\includegraphics[width=0.31\textwidth]{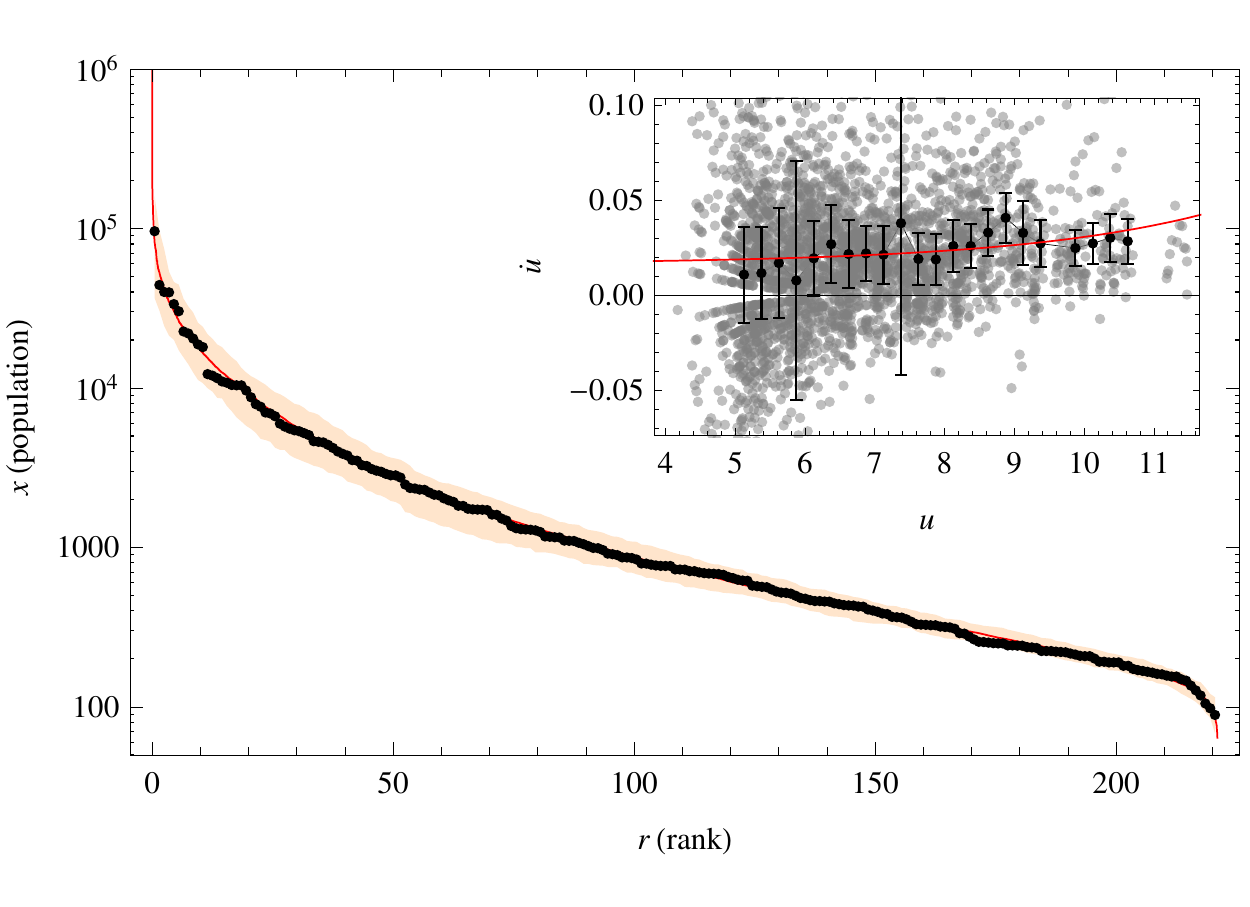}\\
\includegraphics[width=0.31\textwidth]{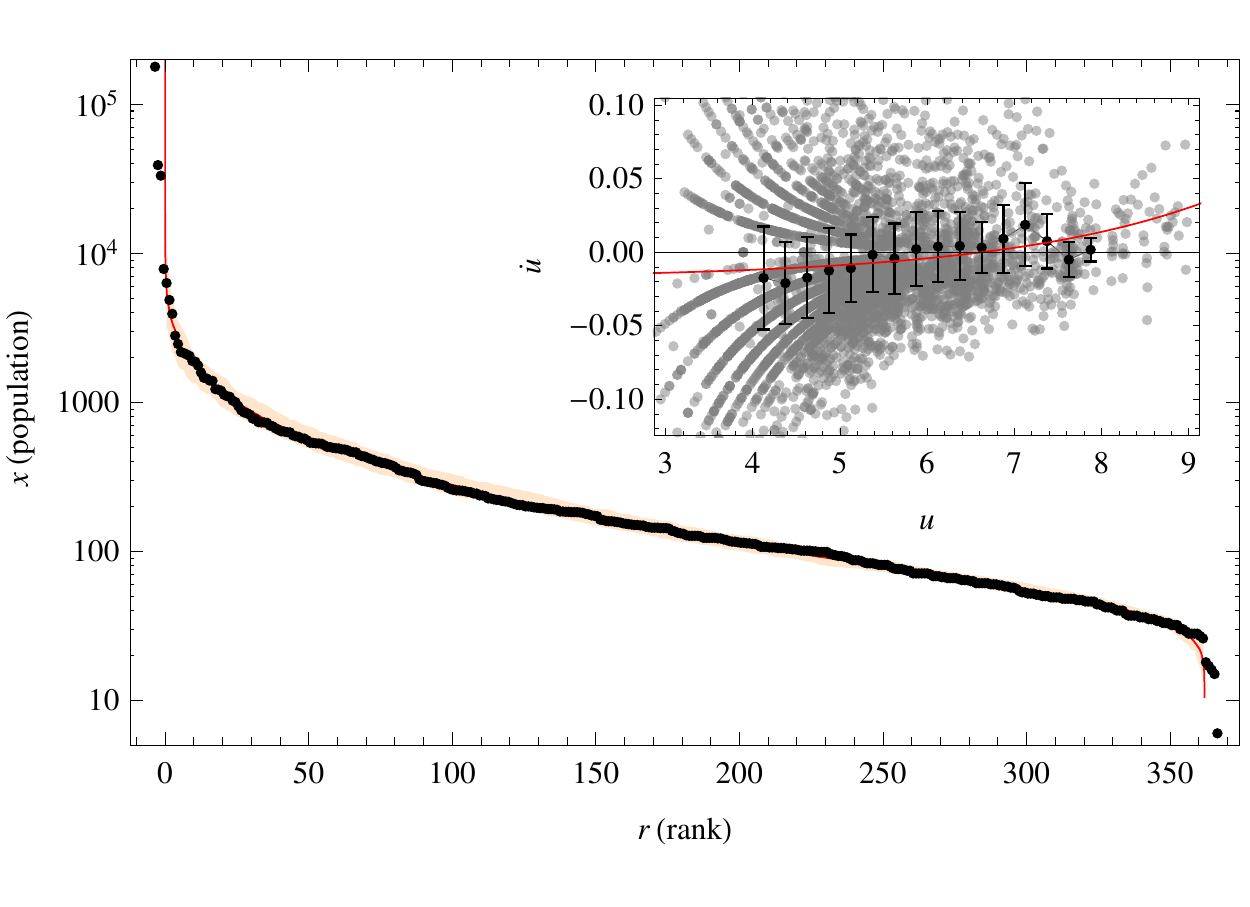}
\includegraphics[width=0.31\textwidth]{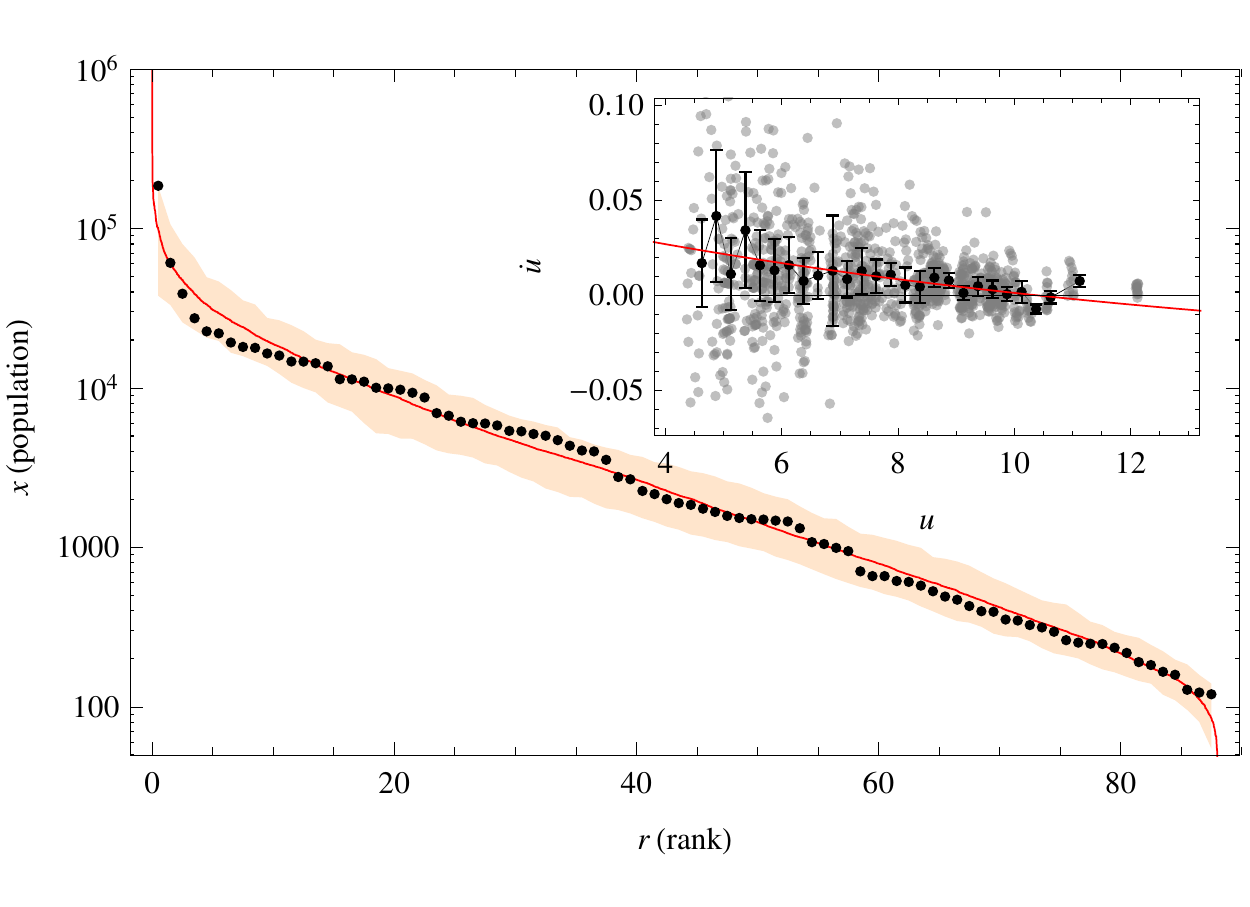}
\includegraphics[width=0.31\textwidth]{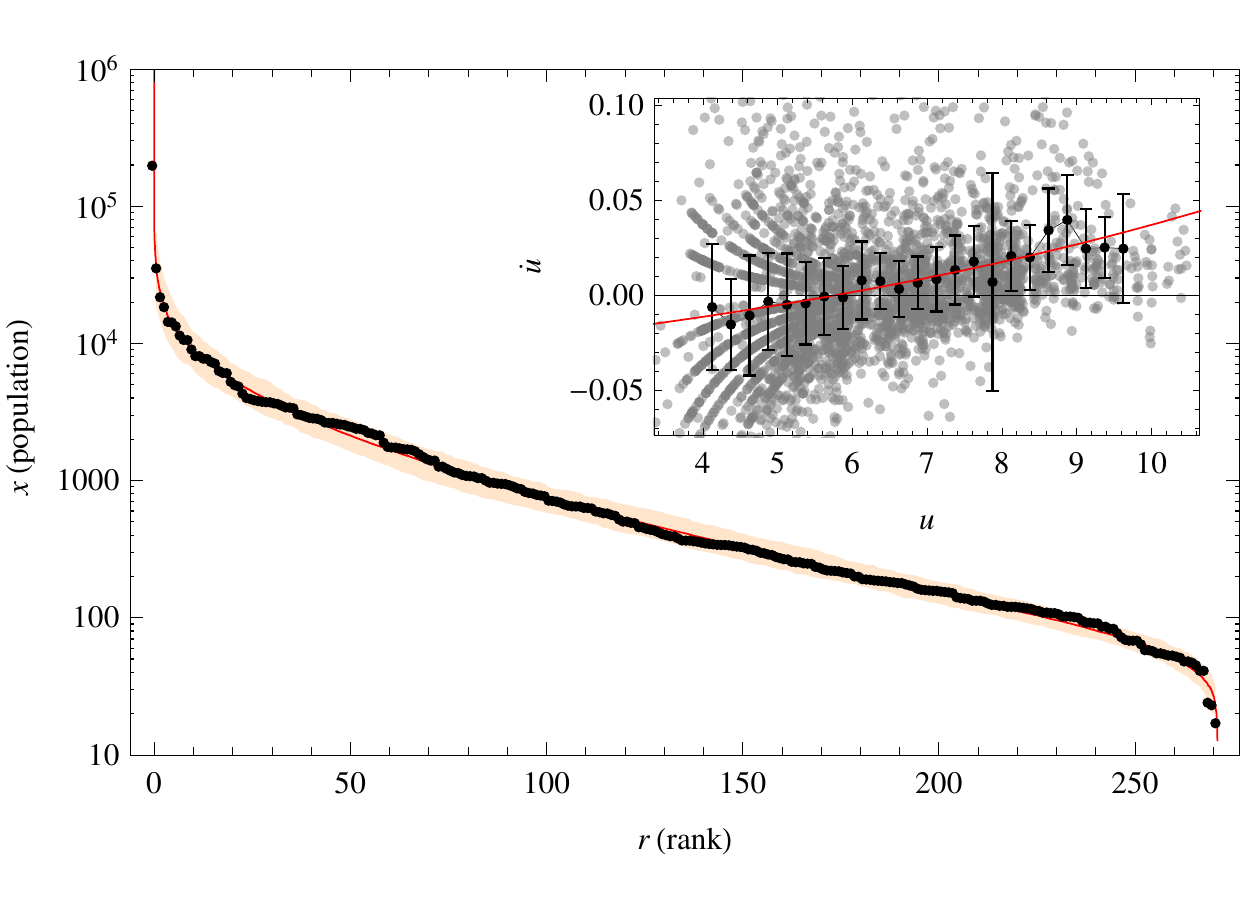}\\
\includegraphics[width=0.31\textwidth]{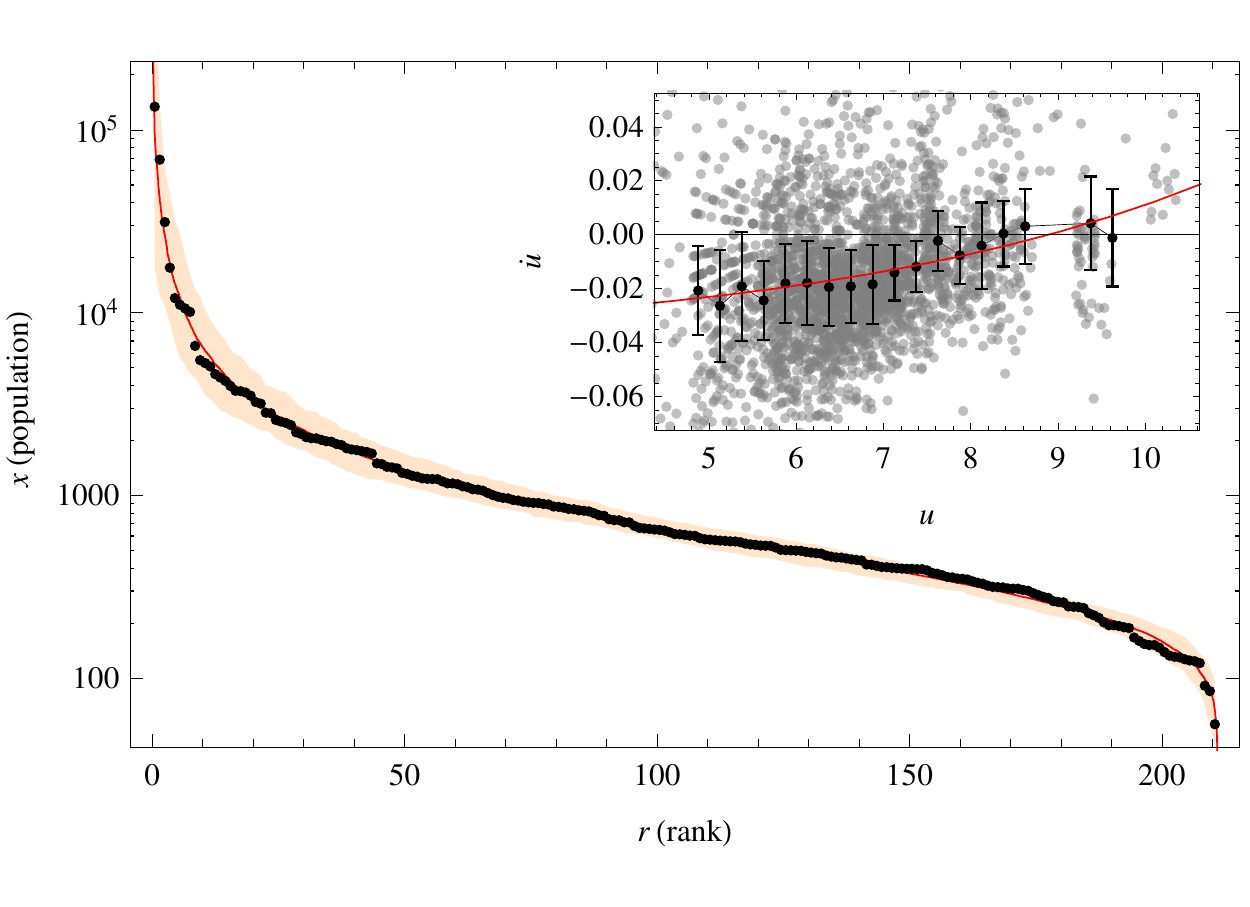}
\includegraphics[width=0.31\textwidth]{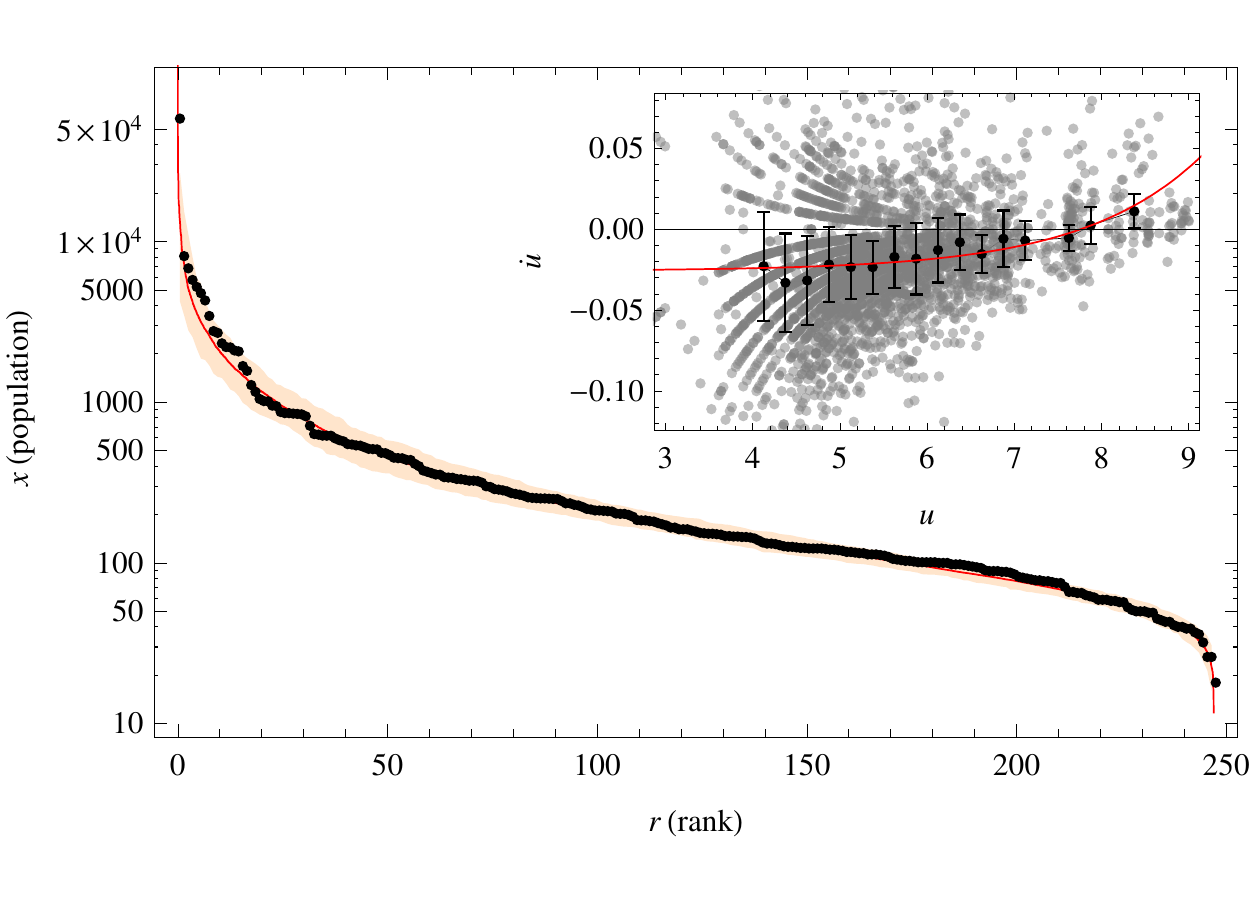}
\includegraphics[width=0.31\textwidth]{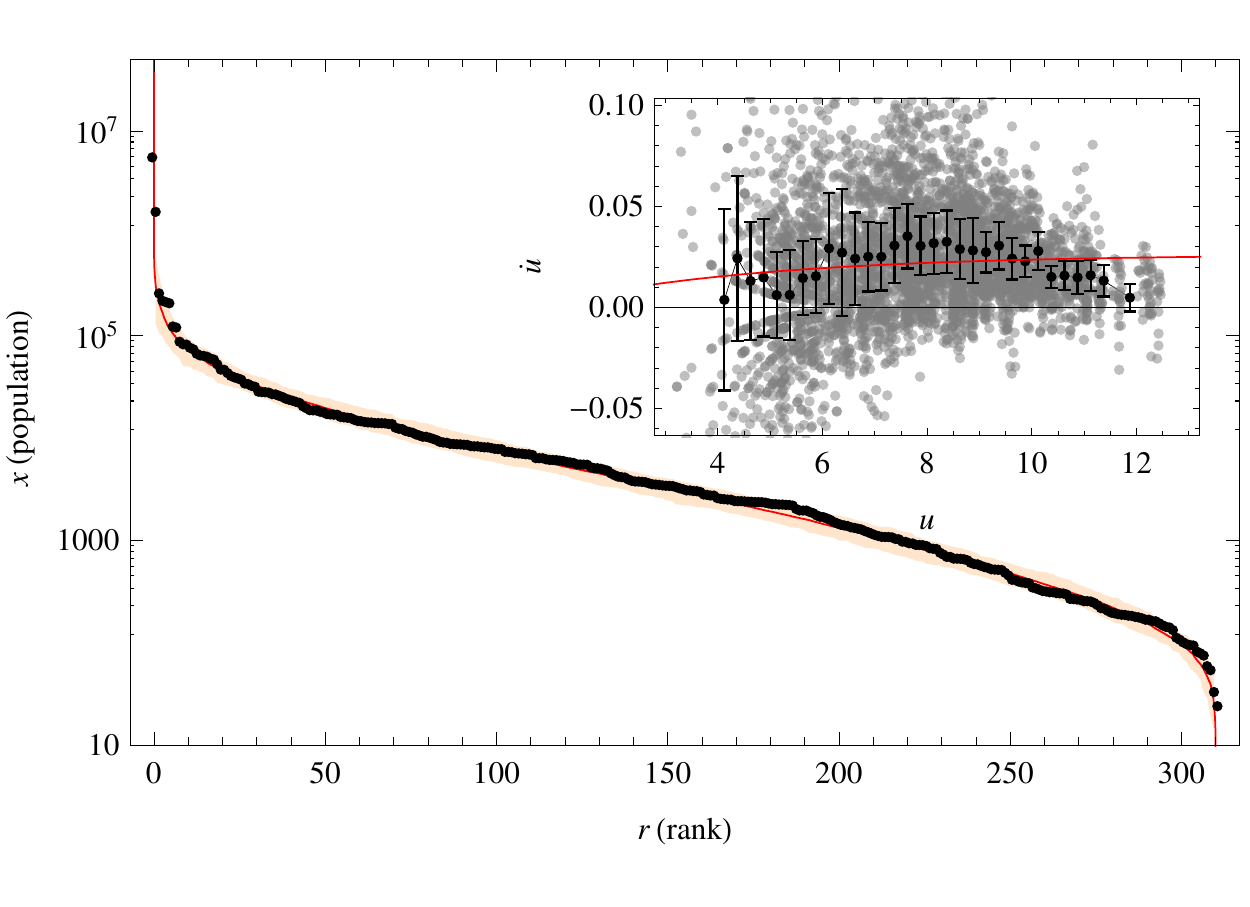}
\caption[]{Rank-distribution for Valladolid, Castell\'on, Girona,
Burgos, Guip\'uzcoa, Navarra, Le\'on, \'Avila, and Barcelona
(year 2010), compared with the pertinent MaxEnt prediction. The
shadowed area represents the 90\% confidence interval for each
$n_c$ value. Insets: relative growth vs. logarithm of the
population, same as inset of top panel in Fig. 1 but with $\Delta
u=0.25$.} \label{fig2}
\end{center}
\end{figure*}

\subsection{Higher order approximation}

\subsubsection{Preliminaries}

\nd    Let us proceed now to tackle a power-law growth by assuming
some kind of  dependence of $\langle\dot{u}(u)\rangle$ on $u$.
Starting from the simplest, linear relation, the idea is to
consider an exponential function and its expansion
\begin{equation}
\dot{u}\simeq a+bu\simeq  k_1+k_qe^{(q-1)u},
\end{equation}
which is equivalent to Eq. (\ref{dynqs}). We   thus
face a ``q-dynamics" with proportional noise, whose
distribution derives from MaxEnt for $q\neq1$ Eq. (\ref{distriqs}).

\nd    We confirm, a posteriori, that this procedure is adequate
up to $q=2$, evidencing   more information  than what one would
get from just a first order relationship. The actual $q-$value can
be determined directly either from the dynamical data or from the
rank-distributions. We can demonstrate via MaxEnt that a
cause-effect relationship exists between the dynamics and the
city-population distribution, if the two kinds of estimations do
match (within standard errors).

\nd    We have estimated the parameters $x_0$, $\Lambda$, $q$, and
$\sigma$ of Eq. (\ref{distriqs}) for each province from the
rank-distribution using the methods described in the Appendix, and
also, now  from the dynamics, the $q$-value for the last 15 years.
We find (mean value $\pm$ standard deviation):
$\log(x_0)\sim5.5\pm1.3$; $\log(\Lambda)\sim-4.3\pm1.6$;
$q\sim1.20\pm0.45$; and $\sigma\sim0.43\pm0.24$.

\subsubsection{Casuistics}

\nd    As a general trend, we can fit  the rank-distribution with
no too many technical complications. In many cases the capital
city has to be excluded, as found in the previous subsection.
Also,  in some other instances, a few of the largest cities have
to be excluded. We have also found `outsiders' for very
low-population centers. More details are described in the
Appendix.

\nd    We encounter  a strong correlation between the number of
municipalities $n_c$ and the minimum size $x_0$ ($R=-0.77$ for
$\log(n_c)$ vs. $\log(x_0)$). In similar fashion as for USA (from
East to West), the area of the Spanish' territories grows form
North to South, due to historical reasons.

\nd    Proportional drift noise, scaled by $\sigma$, has shown to
be negligible in some cases but quite important in others. An low
correlation is found with $\log(x_0)$ ($R=-0.15$), but it is
largely dependent on  $q$ and $\log(\Lambda)$ ($R=0.27$ and
$R=-0.40$). Also, the correlation between $q$ and $\log(\Lambda)$
is rather important ($R=-0.37$).

\nd    The dynamics' fit  presents some preliminary difficulties.
The main one is  high noise  for low city-sizes (the cause can be
found in the last term of Eq. (\ref{clt}), that can not be
neglected). Since the mean value of this noise is zero, we have
found it compelling to estimate $q$ via the `local' mean value of
$\dot{u}$ in Eq. (\ref{dynqs}), expressed as a function of $u$
(see Appendix)
\begin{equation}\label{meand}
\langle\dot{u}\rangle(u) = \langle k_1\rangle+\langle k_q\rangle e^{(q-1)u}.
\end{equation}
Remarkably enough, we systematically find here a better
correlation  than  for the simple linear fitting. In few cases,
even if the variance of $k_q$ is large,  troubles arise when the
value of $\langle k_q\rangle$ is small in relation to  the drift.
In these cases the convergence of the fit falls down to the
generic result $q=1$ and an accurate estimation of $q$ cannot be
achieved from the dynamics.

\subsubsection{Results}

\begin{figure}[t!]
\begin{center}
\includegraphics[width=0.45\textwidth]{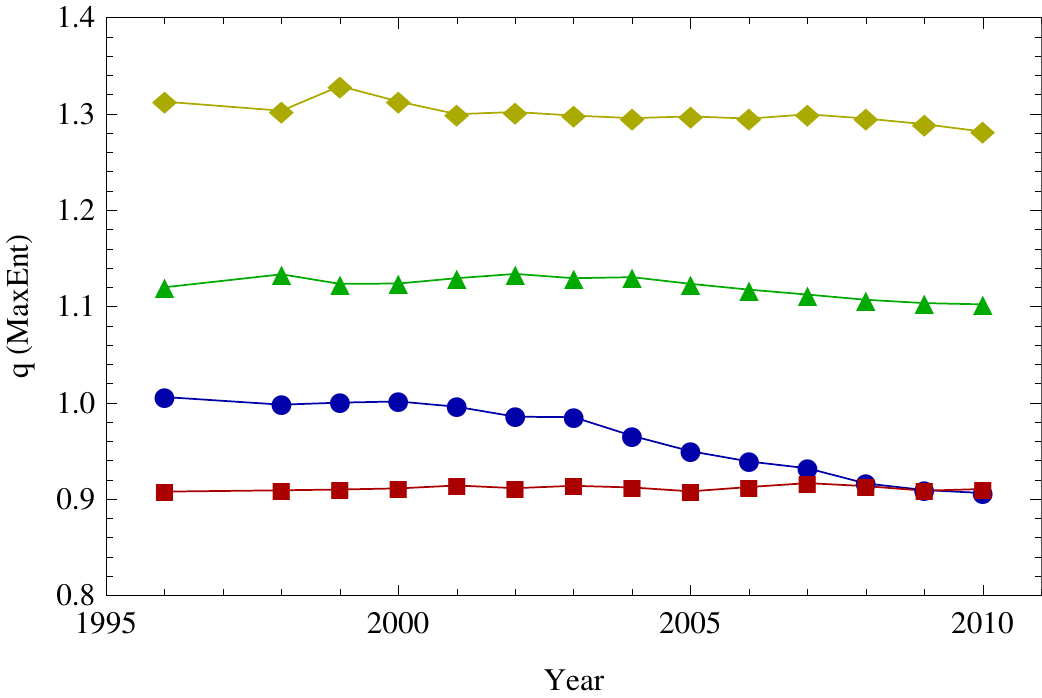}
\caption[]{Evolution of $q$ via MaxEnt distributions for Alicante
(circles), Guip\'uzcoa (squares), Girona (diamonds), and Navarra
(triangles). Alicante is the province with larger $\sigma$
($0.15$) of these four examples, correlated with its larger
time-variation.} \label{fig3}
\end{center}
\end{figure}

\nd    We show in Fig.~\ref{fig2} some interesting examples of
rank-distributions (plots for all provinces are found in the
additional material\cite{AM}, and also the table with all the numerical
values). Most of the provinces follow quite well the analytical
curve,  with very few exceptions (the most dramatic of those
exceptions correspond to  Salamanca, Orense, and Zamora, where  a
few cities account for the main part of the province's total
population and the small villages follow a log-normal
distribution). In general, better fits are obtained for those
provinces for which the distributions' evolution during the last
15 years has been smooth and slow. High rates of change correlate
with high $\sigma$. This rather surprising result tells us that if
the system is able to reach dynamic equilibrium, it converges to
the MaxEnt prediction. We plot in Fig~\ref{fig3} the evolution of
$q$ for some examples, that clearly display the
correlation we are concerned with here.

\begin{figure}[t!]
\begin{center}
\includegraphics[width=0.45\textwidth]{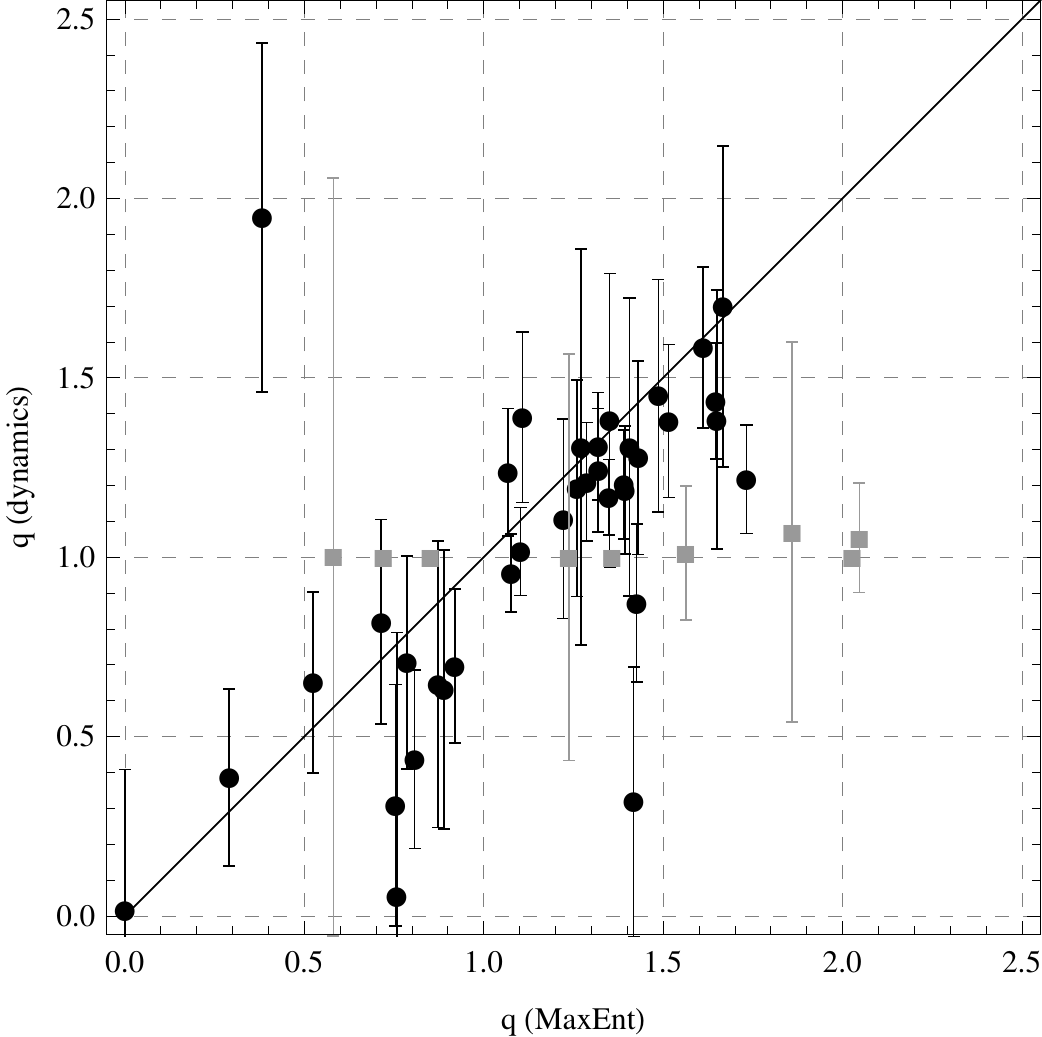}
\caption[]{Comparison of measures of $q$, via the comparison
MaxEnt distribution vs. dynamics. (gray color represents those
cases where the dynamical fit fails, see text).} \label{fig4}
\end{center}
\end{figure}

\nd    We depict in Fig.~\ref{fig4} the comparison between both
independent sources for $q$, MaxEnt vs. dynamics, which represent
the main result of this work (we display in grey the cases where
the dynamical fit fails, near the dynamical value $q=1$). We find
a mean proportionality of $0.85$ with a correlation of $R=0.97$.
In such instances \emph{the dynamical process determines the
rank-distribution of the city-population. The equilibrium
distribution is that found via maximization of the Shannon entropy
in the terms discussed above.}

\section{Further remarks}

\subsection{Origin of the $q$-dynamics}

\nd    In the previous section we have successfully parameterized
the dynamics without  discussing possible underlying mechanisms.
We have encountered two general situations:
\begin{itemize} \item
i) migration from small villages to big cities and \item  ii)
saturation of big cities. \end{itemize} \nd The cases of
\'Avila, Castell\'on, Valladolid and Navarra constitute nice
examples of the first scenario, the most frequent for Spain's
provinces. We usually find for them a value $q>1$. Many
examples share for few of the largest cities the second
possibility, with a relative growth lower than expected.
An explanation for this situation can be found in the
finiteness of the resources to make grow a city, thusly avoiding
arbitrary large growth rates.
Anyhow we have found some examples, as shown for
Guip\'uzcoa, where all the cities follow the second trend and
finding values of $q<1$. These cases
clearly indicate a generalized migration form cities to small towns,
due to some local socioeconomic paradigm.

In other remarkable cases  both situations arise in
the same region, as happens for Madrid and Barcelona. This is a
mark of non-monotonic behavior for $\dot{u}$ with respect $u$,
which in such instance compromises the dynamical fit (shown here
for Barcelona and in the Additional Material for Madrid\cite{AM}).
On the other hand, the case of Girona
deserves some words. Even if it presents a tiny
migration-tendency, the system is very near total equilibrium,
with an excellent fit for the rank-distribution.

\nd    The use of power laws to model both migration and
saturation
 is here of a heuristic nature. We have no evidence of any
`microscopic' mechanism generating them, and the value of $q$ is
merely obtained from empirical observation. Identifying the
mechanism that generates the dynamics constitutes a formidable
challenge. It is worth  mentioning that growth with saturation has
been traditionally modelled using the so-called logistic function,
[or its generalization $Y(t)$] with parameters $\alpha$, $K$, and
$\nu$ that follow a differential equation of the type\cite{wikiglf}
\begin{equation}
\dot{Y}(t) = \alpha\left\{1-\left(Y(t)/K\right)^\nu\right\}Y(t),
\end{equation}
which formally is identical to Eq. (\ref{dynqs}). Accordingly,
migration and growth with saturation generates hyper-exponential
growth, but, as far as we know, a pure theoretical determination
of the $q$-value from the underlying mechanism is still unknown.

\begin{figure}[t]
\begin{center}
\includegraphics[width=0.45\textwidth]{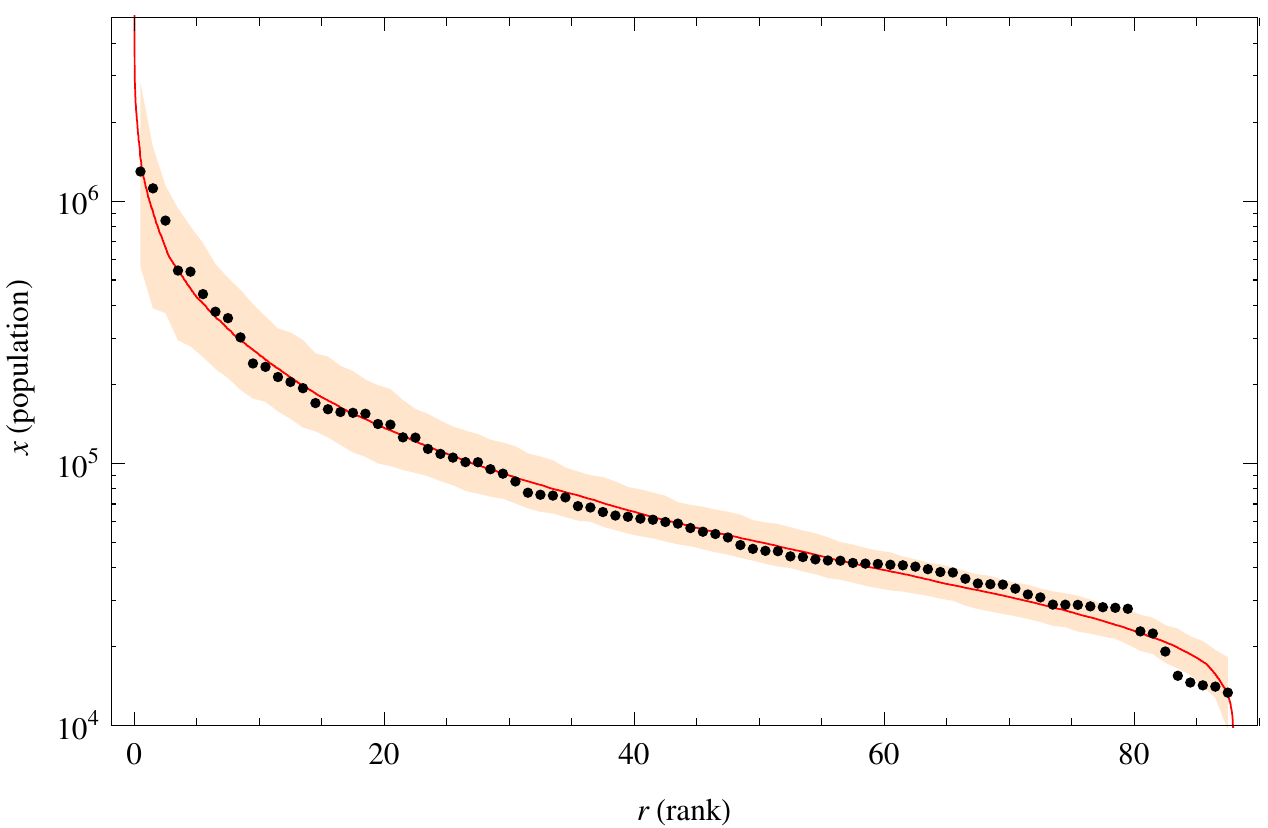}
\includegraphics[width=0.45\textwidth]{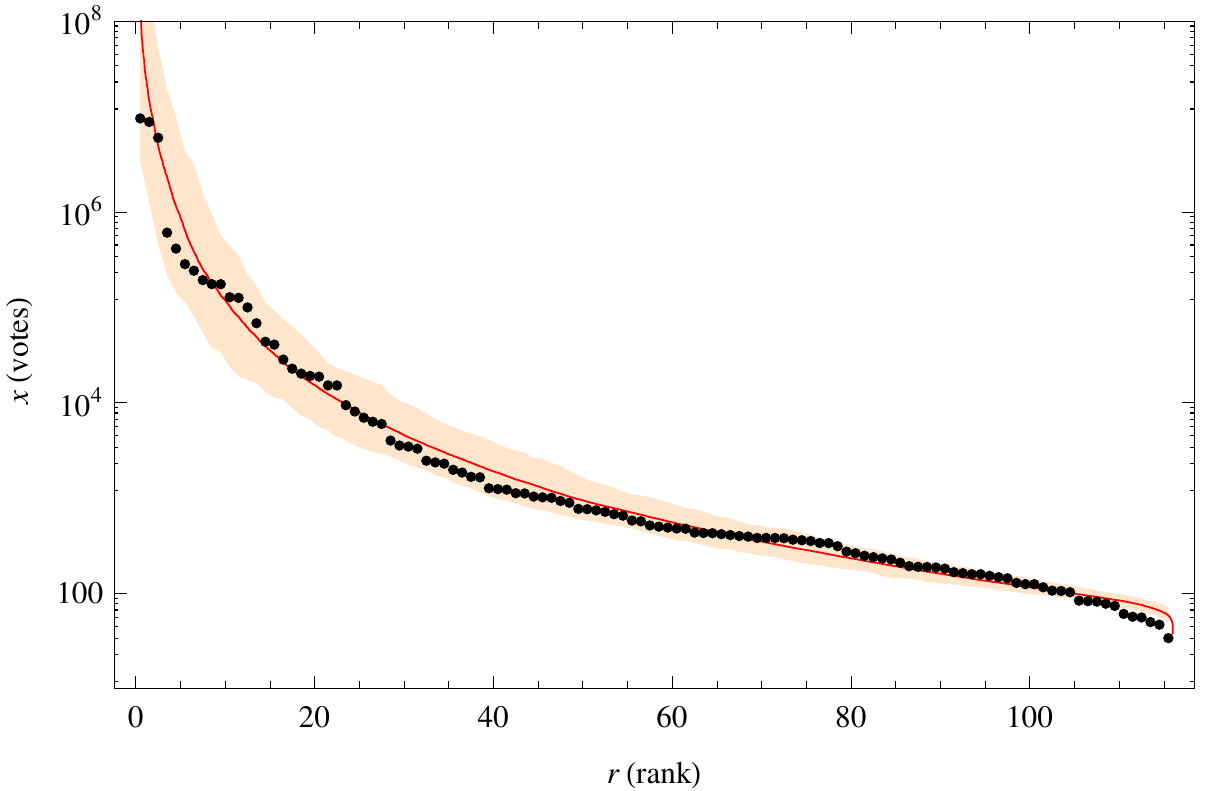}
\caption[]{Same as Fig. 2 for the state of Ohio (USA) in 2010 (top),
and the general electoral results in United Kingdom in 2005 (bottom).}
\label{fig5}
\end{center}
\end{figure}

\subsection{The equation of state}

\nd   We have shown that  the quantities $q$, $\Lambda$, and
$\sigma$ exhibit strong correlations amongst themselves, but not
with $x_0$. MaxEnt gives a relationship between these parameter in
the form of Eq. (\ref{eos}), which might be read as an
\emph{equation of state} that relates all of them. The lack of an
important correlation with $x_0$ is explained because it is a
proportionally constant, being just a shift in the log-scale.
Since we are dealing with scale invariant systems, \emph{the
physics does not depend on scale}, and  the equilibrium
distributions respect this fact. The above correlations may
indicate the appropriateness of using such equation of state in
future city-population research and thermodynamic-oriented
analysis.

\subsection{Beyond Spanish cities' population}

\nd We have also found evidences of MaxEnt equilibrium both in
other countries and in other kind of social systems, as electoral
results, in the present and past years. As an example we plot in
the top panel of Fig.~\ref{fig5} the rank-distribution with
theoretical fit for the state of Ohio ($\log(\Lambda)=-3.7$,
$q=1.62$, $\log(x_0)=10.1$ and $\sigma=0.36$). We also show in the
bottom panel the electoral results for the 2005 UK general
elections ($\log(\Lambda)=-13.5$, $q=1.30$, $\log(x_0)=4.15$ and
$\sigma=0.05$). Deeper understanding of each territory/election,
correlated with the kind of study presented here may help to
better appreciate the effects of regional policies. Work along
such lines is in progress.

\section{Summary}

\nd We have shown that the observed city-population distributions
of the Spanish provinces follow in general the predicted MaxEnt
equilibrium distributions, according to their intrinsic growth
dynamics. \begin{itemize} \item We first have considered a zero-th
order approach to the problem assuming proportional growth, thus
finding that the empirical distributions can be nicely scaled
following the Gamma Scaling Law, derived from the equilibrium
distribution of an scale-free system. \item Secondly, we have
considered hyper-exponential growth with some proportional drift.
We can account in this way for both migration and `natural' growth
of the population, obtaining better fits than in the first
case.\end{itemize} \nd We have checked that the value of the
exponent for the hyper-exponential growth, estimated from the
rank-distributions, is equivalent to the one estimated directly
from the dynamics, which we read as a confirmation of the validity
of our approach.

\section*{}
\nd {\bf ACKNOWLEDGMENT:}  This work was partially supported
by ANR DYNHELIUM (ANR-08-BLAN-0146-01) Toulouse, and the
Projects FQM-2445 and FQM-207 of the Junta de Andaluc\'{\i}a.
AP  acknowledges support from the Senior Grant CEI Bio-Tic GENIL-SPR.

\begin{widetext}
\appendix
\section{Estimation of the MaxEnt distribution parameters from empirical rank-distributions}

\subsection{Estimation of the distribution's 90\% confidence interval}

The confidence levels of the rank-distributions for a given
number $n_c$ are estimated as follows:
\begin{description}
  \item[i)]  A list of $n_c$ random numbers is generated, following the
desired distribution, by inverse transform sampling.
  \item[ii)]  The list is sorted from largest to lowest $n_c$ in order to obtain
the rank-distribution. The list is saved for further use.
  \item[iii)]  A large number of lists is generated following  i) and ii), obtaining
a distribution of numbers for each element of the list.
  \item[iv)]  The 0.95-th and 0.05-th quartiles are obtained from the
distribution  for each element, determining the lower and upper
limits of the 90\% confidence interval.
\end{description}

\subsection{Fit for q-exponential growth}

A first method to estimate the MaxEnt q-distributions' parameters
$\Lambda$, $x_0$, and $q$ for a given empirical data is a direct
fit of the rank-distribution to Eq. (\ref{RDq}). We have used the
\emph{Mathematica} software\cite{math} to this end, by means of the
\texttt{NonlinearModelFit} function using different guesses for
the initial values. We also compare the pertinent results with a
solution obtained from the reproduction of the first three moments
of the distribution, finding, statistically and numerically,
better stability using the logarithmic moments
$\langle(\log(x/x_0))^n\rangle$ instead of
$\langle(x/x_0)^n\rangle$. We have tabulated them as functions of
$q$ and $\Lambda$ in the form
$M_n(q,\Lambda)=\langle[\log(x/x_0)]^n\rangle$, which are defined
via
\begin{equation}
\begin{array}{rl}
\displaystyle \langle\log(x/x_0)\rangle = &\displaystyle
\frac{1}{\Gamma(1-q,\Lambda)}~G^{3~0}_{2~3}\left(\Lambda\left|
\begin{array}{c}
1~1\\
0~0~(1-q)
\end{array}
\right.\right)  \\
\displaystyle \langle[\log(x/x_0)]^2\rangle = &\displaystyle
\frac{2}{\Gamma(1-q,\Lambda)}~G^{4~0}_{3~4}\left(\Lambda\left|
\begin{array}{c}
1~1~1\\
0~0~0~(1-q)
\end{array}
\right.\right)  \\
\displaystyle &\displaystyle \cdots \\
\displaystyle \langle[\log(x/x_0)]^n\rangle = &
\displaystyle
\frac{n!}{\Gamma(1-q,\Lambda)}G^{(n+2)~0}_{(n+1)~(n+2)}\left(\Lambda\left|
\begin{array}{c}
1~\ldots~1\\
0~\ldots~(1-q),
\end{array}
\right.\right)
\end{array}
\end{equation}
where $G$ stands for the so-called Meijer G-functions. The associated
system of equations is
\begin{equation}
M_n(q,\Lambda) = E[\log^n(x/x_0)] = \sum_{m=0}^{n}(-1)^m\left(%
\begin{array}{c}
  n \\
  m \\
\end{array}%
\right)E[\log^{n-m}(x)]\log^m(x_0),
\end{equation}
that we deterministically solve using $n=1,2,3$ via $\Lambda$,
$x_0$, and $q$, with the empirical expected values
$E[\log^n(x)]=\sum_{i=1}^{n_c}\log^n(x_i)/n_c$.

\nd We have also introduced into our equations the proportional
drift parameterized with $\sigma$. Its inclusion in the above
equations can be easily materialized using the additive property
of cumulants for convolutions. Remember that the cumulants of a
probability distribution (PD) are a set of quantities that provide
an alternative to the PD-moments. These, in turn,  determine the
cumulants in the sense that any two probability distributions
whose moments are identical will have identical cumulants as well,
and similarly the cumulants determine the moments. In some cases
theoretical treatments of problems in terms of cumulants are
simpler than those using moments.\cite{wikicum}
Taking into account that for the centered normal distribution
only the second cumulant does not vanish, we face the following
system of equations
\begin{equation}
\sum_{i=0}^{\lfloor n\rfloor}T^n_i\sigma^{2i}M_{n-2i}(q,\Lambda) =
\sum_{m=0}^{n}(-1)^m\left(%
\begin{array}{c}
  n \\
  m \\
\end{array}%
\right)E[\log^{n-m}(x)]\log^m(x_0),
\end{equation}
where $T^n_i$ is the $i$-th element of the $n$-th row of the
triangle of Bessel numbers\cite{series}
$T^n_i=n!/(i!2^i(n-2i)!)$.

\nd The ensuing system of equations is solved using the
\texttt{FindRoot} functionality of \emph{Mathematica} with
different starting points so as to ensure the best possible
result. We accept a set of parameters if the three different
results (fit, and equations' system with/without drift) are
reasonably similar. When that does not happen, we look for
\emph{outsiders}, excluding some points from the tails (largest or
smallest) until reaching a satisfactory convergency of the three
results. The most frequently outsiders-case obtains for  the
capital city. The second case refers to a few small villages with
undersized population. All outsiders are indicated in the
Additional Material\cite{AM}, together with the results of the three
estimations. On the other hand, as commented in the text, we have
found three notorious cases --Salamanca, Orense, and Zamora--
where no satisfactory outcome has been achieved with this
procedure.

\subsection{Fitting the dynamical data}

\nd As commented in the text, we have fitted the mean value
$\langle\dot{u}\rangle(u)$ to Eq. (\ref{meand}) via $\langle
k_1\rangle$, $\langle k_q\rangle$, and $q$. We have employed again
the \texttt{NonlinearModelFit} function, with different guesses
for the initial values. The data set for
$\langle\dot{u}\rangle(u)$ (and also the standard deviation) is
systematically computed in bins of size $\Delta u=0.25$ for all
the provinces. Very few points can be found in some of the bins,
which introduces high numerical error. Accordingly, we include the
bin in the data set if, and only if, a minimal number of points
exists. We have assumed, for all the provinces, that this number
is the 15\% of the bin with the larger number of points. The bins
used in each fit are shown in the inset of the rank-distribution
plots. Anyhow, this filter is not enough in a few cases, for which
a satisfactory result can not be encountered. A $q$-value closed
to 1, albeit ill-defined, is the result.\\

\end{widetext}

\end{document}